\documentstyle[multicol,aps]{revtex}
\begin{document}
\draft
\title{Resonance Patterns of an Antidot Cluster:
From Classical to Quantum Ballistics} 
\author{G. Kirczenow,${^1}$ B. L. Johnson,${^2}$ 
P. J. Kelly,${^3}$ C. Gould,${^{3,4}}$ 
A. S. Sachrajda,${^3}$ Y. Feng${^3}$ and 
A. Delage${^3}$}
\address{${^1}$Department of Physics, Simon Fraser University, 
Burnaby, B.C., Canada  V5A 1S6
\\${^2}$Department of Physics and Astronomy, 
University of New Mexico, 
Albuquerque, NM 87131-1156, USA
\\${^3}$Institute for Microstructural Sciences, 
National Research Council,
Ottawa, Canada K1A 0R6
\\${^{4}}$Universit\'{e} de Sherbrooke, Sherbrooke, Qu\'{e}bec, 
Canada J1K2R1}
\maketitle

\begin{abstract}
We explain the experimentally observed Aharonov-Bohm (AB) 
resonance patterns of an antidot 
cluster by means of quantum and classical
simulations and Feynman path integral theory. 
We demonstrate that the observed behavior of the 
AB period signals the crossover from a
low $B$ regime which can be understood in terms of
electrons following classical orbits to an inherently
quantum high $B$ regime where this classical picture and 
semiclassical theories based on it do not apply.  
\end{abstract}

\begin{multicols}{2}
\section{Introduction}

In recent years the physics of electrons in semiconductor 
systems of reduced dimensionality has been studied extensively at 
high magnetic fields where the quantum mechanics of Landau levels 
gives rise to the quantum Hall effect, and also 
at very low magnetic
fields where many phenomena can be understood in terms of 
electrons following classical ballistic 
trajectories\cite{reviews}. The crossover between these two 
very different regimes is poorly 
understood; identifying and performing relevant experiments 
has been difficult and simple theoretical models that capture 
the essential physics have not been formulated.

A series of experiments carried out recently in the crossover 
regime have yielded intriguing results\cite{Gould}. The device 
studied (see the left inset of 
Fig.\ref{f1}) was a quantum wire 
defined in a two-dimensional electron gas and containing 
a pair of antidots, i.e., regions from which electrons are 
excluded, in close proximity to each other. As a function of 
magnetic field $B$, 
the conductance $G$ of this device showed a dip  
associated with the trapping of electrons in a closed 
orbit encircling the antidot pair\cite{Weiss}.  
Superimposed on this dip were Aharonov-Bohm (AB)
conductance oscillations due to quantum 
interference associated 
with the multiply connected device geometry. Plot (a) 
at the bottom of
Fig.\ref{f1} is an example of such an experimental 
conductance trace. Within experimental 
uncertainty, the period of the AB oscillations was 
{\em independent} of $B$,\cite{Gould}
a phenomenon that has been difficult to understand  
since the area enclosed by a classical cyclotron orbit 
decreases as $B$ increases. One 
might expect this to 
result in an AB period that increases with $B$. \cite{other} 
Another intriguing experimentally observed phenomenon that will be 
reported here is that when the geometry of the antidots is 
varied smoothly by varying the voltage on the gates that define 
them, an additional AB conductance peak can, 
under certain conditions, appear abruptly, 
as if the electronic 
orbit suddenly enclosed an extra quantum of magnetic flux. As 
the gate voltage is varied further, the sequence of 
AB resonances (as a function of $B$) 
soon reforms and becomes regular again. Such dislocations 
in the pattern of AB resonances
are observed when more than one transverse mode is transmitted
through the constrictions between the antidots and the edges
of the quantum wire. The independence of the AB 
period of the value of $B$ is observed both 
in this regime and when only a single mode is transmitted.
 
This is clearly an ideal system for comparing 
classical and quantum behavior. In this article we 
account for its surprising behavior with the help of 
computer simulations and Feynman path integral theory. 
Our quantum simulations 
reproduce the experimentally observed conductance dip and 
AB resonances. We show the latter to 
be due to resonant transmission (as distinct from resonant 
reflection) of electrons. 
We demonstrate that the 
experimentally observed constancy of the AB period is 
an innately quantum phenomenon that
signals the {\em breakdown} of the semiclassical picture 
of electrons following classical trajectories 
that is valid at lower $B$. By contrast, {\em no} 
breakdown of the semiclassical picture   
has been found in previous work on 
large arrays of antidots\cite{tharrays}  
because in those devices\cite{Weiss} the 
constrictions were 
much wider, each transmitting many modes.
Our quantum simulations also reproduce the 
experimentally observed abrupt appearance of additional 
AB resonances as the antidot geometry is 
varied and the subsequent re-emergence of the regular 
sequence of conductance peaks; an explanation 
of this phenomenon is also suggested.   
\section{The Model}

The model used in our quantum simulations is a tight-binding 
Hamiltonian on a square lattice
\end{multicols}
\begin{equation}
H=\sum_{m,n}a^\dagger_{mn}a_{mn}W_{mn}
-t(a^\dagger_{m+1n}a_{mn}e^{in\alpha}
+a^\dagger_{mn}a_{m+1n}e^{-in\alpha}
+a^\dagger_{mn+1}a_{mn}+a^\dagger_{mn}a_{mn+1})
\label{H}
\end{equation}
\begin{multicols}{2}
Here $m$ and $n$ label sites in the $x$ and $y$
directions in Fig.\ref{f1}, $t=\hbar^2 /(2m^*d^2)$, 
$\alpha = eBd^2/\hbar$ (the magnetic phase 
factor $e^{in\alpha}$
is associated with hopping in the $x$ direction since
we use the Landau gauge ${\bf A}=(-By,0,0)$ \cite{ham}), 
${\bf B}$ points in the $z$ direction,
$d$ is the lattice parameter,
$-e$ is the electron charge, $m^*$ is the electron 
effective mass. 
We ignore the small Zeeman splitting 
between spin up and down.
The electron potential energy (the site 
energy $W_{mn}$
in equation (\ref{H}))   
was assumed to be parabolic near the edges of the gates
that define the boundaries of the wire 
and the antidots, and flat elsewhere. 
$W_{mn}$ was modeled 
as a sum of partial electron potential 
energy functions $W_g(u)$ associated with 
the individual
gates $g$, with    
$W_g(u)=E_F (u-a(1+s_g))^2/a^2$ for $u<a(1+s_g)$ and 
$W_g(u)=0$ for $u>a(1+s_g)$. $u$ is the 
distance 
from the edge of the gate. $a$ is a scale of 
length that was chosen
to be $0.05\mu$. $s_g$ is the dimensionless
width of the depleted
region around the gate. (In this work
 $s_g =1$ for both gates that define 
the edges of the quantum wire and 
$s_g \equiv s_d$ for both
gates that define the antidots). 
$E_F=\pi n \hbar^2 /m^*$ is the 
electron Fermi energy measured from the bottom of the band
and $n = 3.47 \times 10^{-15}$ 
m$^{-2}$ is the 2D electron density far from the gates.
As in the experiments, the distance between the outer
gates that define the quantum wire in Fig.1 was $1\mu$, 
each antidot had a lithographic diameter of $0.2\mu$, and  
the spaces between the two antidots and between 
the antidots and outer wire gates were $0.2\mu$. 
In our simulations 100 lattice sites spanned the
$1\mu$ width of the wire. For a typical
value of $B=0.25$T this implies 5 
lattice spacings $d$ per magnetic length 
$(\hbar/(Be))^{1/2}$. Thus although our numerical results
reproduced the experimentally observed behavior of the
device quite well, it should be emphasized that the model
of the system used was a moderately coarse tight binding 
lattice rather than a continuum \cite{mesh}.
\section{Quantum Simulations and Comparison with Experiments}
  
The electron wave functions and the 
multichannel 
quantum transmission and reflection matrices for the above model 
were calculated numerically using a stabilized transfer matrix 
technique\cite{Usuki}. The 
two-terminal conductance $G$ was then obtained from the 
Landauer formula \cite{Imry}
$G={e^2 \over h} {\tt Tr}({\bf tt}^\dagger$). 

The calculated conductance $G$ 
of the device is shown in Fig.\ref{f1} for two 
values of the width 
of the depleted regions around the antidots. 
Plot (b) [(c)] is for $s_d =2.050$ [1.850] for which 
the {\em individual} 
conductances $G_1$ of the two constrictions between the antidots
and the outer gates are both accurately quantized to  
$2e^2/h$ [$4e^2/h$]. I.e., one [two] transverse
modes are transmitted perfectly through each constriction.
The constriction
between the antidots is pinched off.
At the upper and lower ends of the
range of $B$ shown, the conductances of the two parallel
constrictions add (as 
at $B=0$~\cite{Cast}) and the conductance 
of the device is close to $G= 4e^2/h$ [$8e^2/h$].
However, in between $G$ is depressed; this
is the dip in the conductance of the device
that is observed experimentally\cite{Gould} 
due to the trapping of electrons in the 
orbit around the antidots. 
This effect is 
remarkably strong --
near the center of the
dip the calculated conductance falls to little 
more than half of its maximum value.
As in the
experimental data, the calculated $G$ exhibits 
AB oscillations superimposed
on the conductance dip, and as in the data the
pattern of oscillations is much simpler 
when only one mode is transmitted 
through each constriction. (The experimental
trace (a) is in the single mode regime
as is plot (b)).
The conductance dip in the simulations is at
a somewhat higher value of $B$ than in the 
experiment; we did not tune the model parameters 
for a perfect match.
This should be remembered when comparing 
simulations with experiment in Fig.\ref{f2}
and in the top panel of Fig.\ref{f1}. 

A feature
of our numerical results is that  
the AB 
resonances (when they are strong) usually 
take the form of  
sharp peaks and smooth minima in the 
conductance; in the simpler case (b) the peak 
conductance
equals the ideal value $4e^2/h$ for perfect
transmission through both constrictions. 
This means
the AB resonant states 
give rise to {\em resonant
 transmission}. Trace (a) is 
consistent with what would be obtained by 
broadening a series of peaks with sharp maxima
and smooth minima like those in plot (b);
no $kT$ broadening is included in 
plots (b) and (c). 
  
The locations in $B$ of the peaks in the 
calculated conductance $G$ 
are shown in Fig.\ref{f2}(a) for 
various widths $s_d$ of the antidot depletion 
regions.  
The results in Fig.\ref{f2}(a) 
are in good qualitative agreement with the corresponding
experimental data plotted against the antidot
gate voltage $V_g$ in Fig.\ref{f2}(b).
In both the experiment and simulations 
the pattern of conductance maxima is simple and regular
when one mode is transmitted through each of the two
constrictions ($G_1=2e^2/h$). But in the
regime where two modes are perfectly or partly transmitted
there are anomalies -- a 
conductance maximum disappears as $s_d$
increases near
($s_d=1.8$, $B=0.216$T) and as $s_d$ decreases
 near ($s_d=1.97$, $B=0.30$T).
But as $s_d$ is varied further the remaining conductance
peaks form into a periodic pattern once again.
I.e., these anomalies  
resemble dislocations in a crystal.
The same two dislocations are seen at somewhat lower $B$ 
in the experimental data 
when two modes are transmitted through each 
constriction\cite{neg}. A series of
experimental conductance traces around
the dislocation at $B=0.237$T is shown in the inset of 
Fig.\ref{f2}(b).
Note that in the dislocation found experimentally on the 
high [low]
$B$ side of the trapping dip in the conductance, the
conductance peak disappears as the antidot gate 
voltage becomes more [less] negative. This is in 
agreement with the orientation of the corresponding
dislocations
in Fig.\ref{f2}(a) since increasing the depletion 
width $s_d$ in the simulations 
means making $V_g$ more negative. 

Since the dislocations in the pattern of AB 
resonances are found in our simulations
that are based on a model of non-interacting electrons,
it seems unlikely that they 
are due to electron-electron interactions.
However, both experimentally
and in the simulations the dislocations occur when
more than one mode is transmitted through the constrictions.
Indeed more complex dislocation patterns are seen in 
Fig.\ref{f2}(a) for $s_d < 1.7$ where a third mode 
is transmitted. More than one mode penetrating
the constrictions should result in more than 
one trapped electron orbit of the antidot pair
being present. This is consistent with 
the beat-like modulation of the amplitude of 
the AB resonances visible in plot (c) 
of Fig.\ref{f1} (two modes transmitted), 
but absent in plot (b) (one mode penetrates).
Our numerical study of the 
transmission probabilities associated with 
{\em individual} scattering channels that contribute
to the total conductance of the device 
showed that
when two modes are transmitted through the constrictions 
the resonant structures in the different channels 
fall into two groups. The peaks in
the transmission belonging to these two 
groups move out
of phase with each other in the vicinity of the
dislocations and cancel there resulting in the 
disappearance of a conductance peak.  
Thus it appears that destructive interference of
out of phase resonances
associated with at least two different orbits is 
responsible for the dislocations. 

The spacing of 
successive conductance peaks in $B$ 
(the AB period $\Delta B$) 
found in the quantum
simulations is shown by the large open [full]
circles at the top of Fig.\ref{f1} for the same
$s_d$ as in conductance plot (b) [(c)].
$\Delta B$ decreases with increasing $B$ at low $B$
but is 
{\em stationary} at the conductance 
dip. (The anomaly in $\Delta B$ at $B=2.8$T for 
case (c) is due to a dislocation whose
core is at ($s_d=1.65, B=3.0$T) 
in Fig.\ref{f2}(a)). Since $\Delta B$ 
could only be measured in a relatively
narrow range of $B$ at the conductance dip, the fact that
the simulations show $\Delta B$ to be stationary there
accounts for the independence of $\Delta B$ of $B$ that was
observed by Gould {\em et al.}\cite{Gould}. The small 
symbols in the top panel of Fig.\ref{f1} show 
the measured $\Delta B$ for several  
values of $V_g$. The
data shows some scatter but there is overall 
agreement with
the quantum simulations -- the measured
$\Delta B$ also decreases with increasing $B$ at low
$B$ and levels out at the conductance dip. Notice that
overall $\Delta B$ is insensitive to $s_d$ in both
Fig.\ref{f1} and Fig.\ref{f2}(a), consistent with
the insensitivity to the antidot gate
voltage found experimentally\cite{Gould}.
\section{Analysis: Quantum Simulations vs. Classical 
Simulations and Feynman Path Integral Theory}

The fact that $\Delta B$ 
{\em decreases} 
with increasing $B$ at low $B$ may at first seem 
surprising since the area enclosed 
by a cyclotron orbit in free space 
decreases with increasing $B$. However, at low
$B$ the diameter of a cyclotron orbit
is larger than the distance between
the gates defining the edges of the wire, so that 
a closed {\em circular} orbit enclosing the 
antidots can not occur. But 
the constrictions between the the gates
and the antidots are asymmetric; one side is curved,
the other straight -- See left 
inset, Fig.\ref{f1}.
This results in the electron
wave, as it emerges from the constriction, 
being collimated
at an angle of about $\pi/6$ relative to the edge of the
straight gate\cite{Akis} so that a closed 
{\em non}-circular quantum 
orbit of the antidots can form.
 
A similar collimation effect also occurs 
classically\cite{Akis}, making closed classical low $B$
orbits possible. We have performed classical 
calculations of such closed orbits for the same model
electron potential as was used in our
quantum simulations and calculated the AB phase 
associated with these orbits using the
semiclassical result
$\hbar \phi_{AB}^{S.C.}=
\oint m^* {\bf v}.d{\bf l} - eA_{e} B$. Here 
$A_{e}$ is the area enclosed by the orbit. 
The semiclassical AB period thus obtained for a
closed classical orbit is shown by
the line in the top panel 
of Fig.\ref{f1}; the
semiclassical AB period decreases
with increasing $B$ as in the quantum simulations
but does {\em not} become 
stationary\cite{classical}. 
Since the same electron potential energy function was
used in both the quantum and classical simulations,
the fact that the stationary AB period was only
found in the quantum simulations is significant:
It implies that 
the constant
AB period observed experimentally is an inherently
quantum phenomenon that
signals the breakdown of the semiclassical picture 
of electrons following classical trajectories 
that is valid at lower $B$. 
  
One can understand the effect 
of the confinement
of the electron orbit by the edges of the 
quantum wire on the AB period at low $B$ 
by considering
a simplified model of the device
and a class of Feynman paths 
depicted in the upper right
inset of Fig.\ref{f1}. The edges of the wire
are represented by infinite hard walls
and the antidots by a narrow infinite 
barrier. In the symmetric gauge 
$A=(-By/2,Bx/2,0)$, 
we parameterize the Feynman path by
$x=k(v/\omega)\cos(\omega t)$, 
$y=(v/\omega)\sin(\omega t)$,
where $t$ is the time, $v$ is the Fermi velocity
and $\omega = v/r$. $r$ is half of the distance
between the constrictions.
$k$ is the parameter that defines the shape of the 
particular Feynman path.
The action for a closed path is then given by 
$S_k = \oint L dt =  
Tm^* v^2 (k^2-2k\omega_c/\omega+1)/4$ where 
$T =2 \pi /\omega$
and $\omega_c=e|B|/m^*$. The action is stationary with 
respect to the shape parameter $k$ for 
$k=\hat{k}=\omega_c/\omega$. Thus the stationary Feynman
path is circular as expected when $\omega_c =\omega$,
i.e., when the cyclotron radius equals $r$. For lower
$B$, $\omega_c <\omega$ and therefore
$\hat{k}<1$ and the stationary Feynman
path is an ellipse whose area {\em increases}
as the $B$ increases. This suggests
that the AB period should decrease
with increasing $B$, qualitatively in agreement
with our low $B$ simulation results. On closer
inspection, this 
turns out to be correct -- The propagator 
$U({\bf r},t,{\bf r},t+T)$ for the set of Feynman
paths enclosing the barrier is dominated by 
$e^{iS_{\hat{k}} / \hbar }$, and therefore
the AB phase is given (up to a constant)
by $\phi_{AB} \approx S_{\hat{k}} / \hbar = 
Tm^* v^2 (1-\omega_c^2/\omega^2)/(4\hbar)$. The 
AB period is then given by 
$2\pi=|\Delta \phi_{AB} |\approx \Delta B~ |d\phi_{AB}/dB|$
or $\Delta B \approx 4\hbar m^* \pi \omega^2/(BTv^2e^2)$.
I.e., $\Delta B$ decreases as $B$ increases. More 
general path integral calculations that confirm  
this qualitative result will be presented 
elsewhere.  

At higher $B$ the cyclotron diameter
becomes smaller than the width of the quantum wire,
so that the confinement of the AB orbit 
by edges of the quantum wire should be less
important. Thus at higher $B$ one should
expect the area of the quantum AB orbit to 
shrink and the AB period to {\em increase}
with increasing $B$. The crossover to this higher
$B$ regime should occur when the cyclotron diameter
is equal to the distance between the two constrictions
at the edges of the device, i.e., within the dip in the
conductance of the device that is due to the 
trapping of electrons in the orbit encircling 
the antidots. This explains why the 
AB period in our quantum simulations begins to
increase with increasing $B$ at the high $B$ edge
of the resistance peak, as can be seen in 
Fig.\ref{f1}. 
Thus $\Delta B$
was effectively independent of $B$
in the experimental data of 
Gould {\em et al.}\cite{Gould} because that data
was taken at the crossover from the low
$B$ regime to the high $B$ regime, which coincides
with the trapping dip in conductance\cite{clarify}.

As a further test of this explanation we carried
out quantum simulations of transport in a similar device
in which the straight edges of the quantum wire 
are further from the constrictions and should
have less effect on the AB orbit at
the lower values of $B$. This device is depicted
in the lower right inset of Fig.\ref{f1}.
As expected from the above argument, 
in our simulations of this device
the AB period 
increased with increasing $B$ throughout the range
of $B$ in which AB conductance oscillations
were present. Experiments 
by Hirayama and Saku \cite{Hirayama} on such a 
device also found the AB period to 
increase with increasing $B$, lending further
support to the above theory.
\section{Conclusions}

In conclusion, we have shown that the observed
Aharonov-Bohm resonance patterns of an antidot 
cluster can be accounted for by quantum simulations
and have demonstrated that the previously  
unexplained constant period of the
resonances signals the crossover from a classical 
ballistic regime at low magnetic fields to an 
inherently quantum regime at higher fields. The 
crossover from classical to quantum ballistics
should be an interesting new field of study.

We thank Richard Akis for a helpful suggestion.
This work was supported by NSERC of Canada.

\end{multicols}
\begin{figure}
\caption{Two-terminal conductance $G$ and 
AB period $\Delta B$
vs. $B$ for device in 
left inset. Plot (a) [(b)]: experiment [quantum 
simulation] with
one mode transmitted through each constriction.
Plot (c): quantum simulation with two modes
transmitted. Large open [full] circles: $\Delta B$
for case (b) [(c)]. Small open 
[full] symbols: experimental $\Delta B$ 
in one [two] mode regime. Solid line in top panel: 
semiclassical $\Delta B$ vs. $B$ for closed classical 
orbit of antidot pair; $s_d=2.0$.  
Left inset:
Schematic of quantum wire with two 
antidots. Black and shaded 
regions are depleted of electrons. Upper right inset: 
Simplified model used in analytic path integral theory.
Lower right inset: Device in which the straight edges of the
wire are further from the constrictions.}
\label{f1}
\end{figure}
\begin{figure}
\caption{Positions in $B$ of conductance maxima 
obtained from 
(a) quantum simulations vs. width $s_d$ of the 
antidot depletion regions and (b) measurements vs.
antidot gate voltage $V_g$. Quantized plateaus in 
conductance $G_1$ of {\em individual} constrictions 
are indicated. Curves are guides to the eye. Inset: Measured
conductance $G$ [for various $V_g$ from -1.16V (top
trace) to -1.32V] vs. $B$ (from 0.21T to 0.26T)
around dislocation at $V_g=-1.22$V, $B=0.237$T.
Traces offset for clarity.}
\label{f2}
\end{figure}

\begin{references}
\bibitem{mesh} Performing extensive quantum simulations
such as those reported here using significantly finer lattices 
is not practical at present because of the 
limitations of computers.
\bibitem{classical}We found several different closed 
classical orbits. The AB period behaved similarly 
for all of them.
\bibitem{clarify}$\Delta B$ is stationary at the crossover
and is thus observed experimentally to be constant 
there because the crossover is from a low $B$ regime where 
$\Delta B$ decreases with increasing $B$ to a high $B$
regime where $\Delta B$ increases with increasing $B$,
as is discussed in detail above.
\end{references}
\end{document}